\documentstyle[aps,preprint]{revtex}
\begin{document}
\draft
\tightenlines
\title{Spontaneous formation and stability of small GaP fullerenes}

\author{V. Tozzini}

\address{
Istituto Nazionale per la Fisica della Materia and
Scuola Normale Superiore, 
Piazza dei Cavalieri, 7 
I-56126 Pisa, Italy }

\author{  F. Buda}

\address{
Department of Theoretical Chemistry, Vrije Universiteit, 
  De Boelelaan 1083, NL-1081 HV Amsterdam, The Netherlands }

\author{A. Fasolino}

\address{
Research Institute for Materials, 
Institute of Theoretical Physics, 
University of Nijmegen, Toernooiveld,
NL-6525ED Nijmegen, The Netherlands }

\maketitle

\begin{abstract}

We report the spontaneous formation of a GaP fullerene cage 
in ab-initio Molecular Dynamics simulations starting from a
bulk fragment. 
A systematic 
study of the geometric and  electronic properties of neutral 
and ionized GaP clusters suggests the stability of 
hetero-fullerenes formed by a compound with zincblend bulk structure. 
We find that GaP fullerenes up to 28 atoms have high symmetry, 
closed electronic shells,  
large HOMO-LUMO energy gaps and do not dissociate when ionized.
We compare our results for GaP with those obtained by other groups for
the corresponding BN clusters.

\end{abstract}
%\begin{multicols}{2}

The discovery of carbon fullerenes and nanotubes\cite{scopertaful,scopertanan}
has opened a completely new 
field at the borderline between chemistry and physics leading to many
new phenomena and applications.

Up to now, most efforts to identify fullerenes based on other elements have 
focused on  BN  which is the most similar to carbon and  exists in nature 
in the hexagonal(graphite-like) structure 
~\cite{BNexp,Parilla,Seifert,Toftlund,Blase,Sun,Alexandre,Fowler}. However, 
the (nested-)cages and wires found for 
this material\cite{BNexp,Parilla} do not resemble any of the small preferred 
structures of the carbon fullerene family,
particularly due to the absence of the characteristic pentagonal rings.
Besides, nanotubes based on other layered materials, such as 
GaSe~\cite{Chadi} and black-phosphorus~\cite{hernandez},
have been theoretically predicted to be stable.  
On the basis of density functional calculations
it has also been proposed
that GaN nanotubes could be synthesised by using carbon nanotubes 
as a nucleation seed\cite{seung}.

One intriguing question is whether fullerene cages could be realized 
in typical semiconductors of the III-V family, like GaAs, 
InSb or GaP, which do not possess a graphite-like bulk structure.
These materials are not considered
as good candidates for hollow structures 
since $\pi$ bonding should be less effective in these larger atoms of 
higher rows of the periodic table than in the first one\cite{Seifert}.

In this letter we show, by means of ab-initio Car-Parrinello Molecular 
Dynamics\cite{CP}, that a small
GaP bulk fragment spontaneously organizes in a cage formed by a 
different number of atoms of the two elements arranged 
as in carbon fullerenes. We discuss the geometric and  electronic structure
of GaP cages with either 
the same or a different number of atoms of the two species.
Our results strongly suggest that small GaP fullerenes could be stable, 
since they have high symmetry, 
closed electronic shells,  
large HOMO-LUMO energy gaps and do not dissociate when ionized.
We give quantitative estimates of the relative stability of 
cages formed either by hexagons and pentagons as in carbon fullerenes or by 
hexagons and squares as proposed for BN\cite{Seifert,Toftlund,Blase,Sun,Alexandre}. 

Our results are obtained by the 
Car-Parrinello  approach\cite{CP} using a Density Functional 
in the Generalized Gradient Approximation proposed by 
Becke and Perdew\cite{Becke,Perdew}. This approximation 
reproduces the experimental cohesive energy of typical 
bulk semiconductors within $\approx 5 \%$ and underestimates the
valence to conduction band excitation energies \cite{Ortiz}. 
We use nonlocal norm-conserving first-principles pseudopotentials\cite{Gonze} 
and expand the single particle wavefunctions on a 
plane wave basis set with a cut-off of 12 Rydberg. We use a periodically 
repeated cubic 
simulation box of 24 \AA\ side, so that periodic images are 
at least 14 \AA\ apart. We have verified that this size is large enough 
to describe isolated clusters. 
The electronic optimization and
structural relaxation have been performed using damped second order dynamics
with electronic mass preconditioning scheme~\cite{tassone,PRB2}. 
We use throughout an integration time step of 8 a.u.
The symmetry of the equilibrium structure is not biased but 
it is reached spontaneously 
during the geometry optimization starting from the corresponding regular
polyhedron. 

The process of formation of the 
fullerene cage with 28 atoms from a larger bulk-like cluster of 41 atoms 
(Ga$_{28}$P$_{13}$) is 
shown in Fig.1, with the help of three snapshots taken 
during the structural energy
minimization which leads to the appearance of the
Ga$_{16}$P$_{12}$ fullerene cage~\cite{nota21}. 
The cage has 12 pentagons and 6 hexagons and 
T$_d$ symmetry as C$_{28}$. 
An analysis of the charge distribution shows 108 valence electrons 
on the bonded cage, 
exactly the number which corresponds to the neutral Ga$_{16}$P$_{12}$ 
cluster\cite{nota}. Finally in Fig.1d we show the equilibrium structure
of the neutral Ga$_{16}$P$_{12}$ cluster alone. 

The observed spontaneous formation of a Ga$_{16}$P$_{12}$ cage with 
pentagons is surprising since, in the case of 
BN\cite{Seifert,Toftlund,Blase,Sun}, (deformed) squares are found to be 
energetically much more favorable. 
For B$_{12}$N$_{12}$~ \cite{Toftlund}, 
there is an energy difference of 9~eV between the cage with 
pentagons and the one with squares in favor of the 
latter which contains only heteropolar bonds and 
is favored for a material 
composed by atoms with very different electronegativity as B and N. 
Therefore, most studies have considered cages B$_n$N$_n$
formed by hexagons closed by square rings\cite{Seifert,Toftlund,Sun,Alexandre}.
Very recently, 
Fowler et al.\cite{Fowler} have pointed out 
that, among the cages with pentagons, those with one species in excess of 
4 atoms (B$_n$N$_{n+4}$) minimize the number of homopolar bonds.
It is remarkable that the cage Ga$_{16}$P$_{12}$ 
which spontaneously appear in our simulation falls into this class.

We have studied the equilibrium structure and 
electronic states of clusters with 20 and 28 atoms of the type 
III$_n$V$_{n\pm4}$ , namely  Ga$_{12}$P$_{8}$, Ga$_{8}$P$_{12}$, 
Ga$_{16}$P$_{12}$ and Ga$_{12}$P$_{16}$, and compared them to clusters 
with the same number of 
III and V atoms, namely Ga$_{10}$P$_{10}$ and Ga$_{12}$P$_{12}$, the latter
in the two isomers\cite{Toftlund} 
with hexagons closed either by five- or four-sided faces. 
The minimum energy structures of Ga$_{16}$P$_{12}$ and Ga$_{12}$P$_{16}$ 
are found to have  T$_d$ symmetry, whereas those of 
Ga$_{12}$P$_{8}$ and Ga$_{8}$P$_{12}$ have T$_h$ symmetry. Among the clusters 
Ga$_n$P$_n$,  the Ga$_{12}$P$_{12}$ with 4-membered rings belongs to T$_h$ 
whereas those with pentagons present very large distortions around the lower
C$_{3v}$ symmetry. 
The structural parameters  of the cages 
belonging either to T$_h$ or to T$_d$ are given in Table I. 
Hexagons are found with 
alternating angles of $88^0-105^0$ and 
$126^0-134^0$, pentagons  
with angles of  $85^0-92^0$, $ 114^0-126^0$,  
$\sim 100^0$, $\sim 110^0$
and 
deformed squares with angles $ 75^0$ and $98^0$.  
Ga-P distances are in general shorter than in bulk compounds due to 
predominant $sp^2$ bonding.
The radial distance  $r$ from the center 
of the cluster given in Table I indicates a tendency of the anion to 
occupy positions at larger distances from the center than the cation, as
found for ultrasmall cluster\cite{Andreoni,PRB1}. 

As in the case of carbon and BN fullerenes, the GaP clusters would represent 
metastable states with respect to the bulk equilibrium structure.  
Therefore only experimental observation can establish with certainty 
their existence. Nevertheless, there are a few quantities which are used 
in the literature as indicators of stability. We support our prediction 
for the stability of the examined GaP clusters by using the following 
indicators: i) closed electronic shells and large energy gaps; ii)
cohesive energy; iii) thermal stability;  iv) stability of the ionized 
clusters.

The first indicator of chemical stability is the energy gap between 
the highest occupied and lowest unoccupied molecular orbitals (HOMO and LUMO).
In carbon fullerenes a correlation between this energy 
and the observed fullerenes has been experimentally verified\cite{egap}. 
In Table II we give the calculated  HOMO-LUMO energy gap and the cohesive 
energy for all the clusters studied.  
Among the cages  with pentagons the highest energy gaps
are for cluster with P in excess of 4, a  
composition which has been suggested to be favorable also for BN\cite{Fowler}.
However, a  very large gap  is also found for the 
Ga$_{12}$P$_{12}$ with squares. 

A comparison of the binding energies per atom between the GaP cages and
the zincblend bulk phase of this material 
is possible only for the clusters with the same number of Ga and P atoms.
From the results of Table II, we find that the cohesive energies per atom 
for Ga$_{12}$P$_{12}$ with squares, Ga$_{12}$P$_{12}$ with pentagons,
and Ga$_{10}$P$_{10}$
are about 10$\%$ lower than in the bulk. 
This result is very close to that found for BN and carbon fullerenes
of the same size~\cite{Seifert}.

We have studied the thermal stability of 
two clusters with very different energy gaps, namely 
Ga$_{12}$P$_{8}$ and Ga$_{8}$P$_{12}$ (see Table II). 
For both clusters we 
have performed two annealing cycles of about 3 ps, 
up to 1500 K and up to 2000 K.
The system is heated with a rate of
2$\times$10$^{15}$K/s, then equilibrated for 
one ps at the highest temperature, and finally cooled down with 
the same temperature change rate.
For Ga$_{8}$P$_{12}$ no bond breaking or structural rearrangements occur 
in both cycles and the structure 
comes back to the same minimum energy configuration when the temperature 
is lowered. This is also the case for Ga$_{12}$P$_{8}$ in the
annealing up to 1500 K, whereas at 2000 K some structural 
rearrangement takes place leading  to a distorted structure with higher 
energy when cooled down.
These results indicate that the thermal stability is correlated with
the width of the energy gap.

Mass spectrometry experiments use the difference in mass-to-charge ratio 
of ionized atoms or clusters to select them. 
Therefore one basic requirement for the possible detection of such
clusters is that they remain stable also when ionized.
We have investigated 
the stability of some positively ionized clusters, [Ga$_8$P$_{12}]^+$, 
[Ga$_{12}$P$_{16}]^+$ and [Ga$_{16}$P$_{12}]^+$. 
We have included an uniform charge background in order to have an
overall neutral system in the supercell calculation.
The electronic 
structure remains almost unaffected and degeneracies are broken by 
negligible amounts in the order of hundredths of eV. 
Only minor structural distortions occur upon ionization. 
In particular, the six equivalent P-P (Ga-Ga) bond lengths
split into three different classes. Remarkably, during a molecular
dynamics run for [Ga$_8$P$_{12}]^+$ 
we observe a dynamical exchange between these three 
classes of bond lengths with each other. 
This effect produces features in the low frequency 
vibrational spectrum in the range $30-120$ cm$^{-1}$ which might be detected
by infrared multiphoton ionization spectra\cite{Deniz}.

It is interesting to compare cages 
closed either by 4- or 5-membered rings. 
As already mentioned, such a comparison has been done for 
B$_{12}$N$_{12}$ in Ref.\onlinecite{Toftlund}. However there are no results 
comparing clusters with equal number of atoms 
III$_n$V$_{n}$ with square rings to 
the more favorable structures with pentagons 
of the type III$_n$V$_{n+4}$. The authors who have proposed the latter 
stoichiometry\cite{Fowler}, in fact, do not give a comparison to cages
with square rings. Although
it is impossible to compare directly the cohesive energy of structures 
with a different number of atoms of each species, we are in a position to give 
an estimate of the bond energy for the two types of cages.
Given the number of each type of bond in all the
structures with pentagons studied so far and the values of the 
total cohesive energy,
we estimate by a best fit the following energies per bond: E$_{Ga-P}$=
-2.593 eV,  E$_{Ga-Ga}$=-1.133 eV,  E$_{P-P}$=-2.349 eV\cite{emsley}. 
As shown in Table II,
these values yield the correct cohesive energy with a relative error 
of 0.4\% at most for all cluster with pentagons, whereas overestimate
the cohesive energy of the cluster Ga$_{12}$P$_{12}$ with squares. 
In this cluster, in fact, there are only Ga-P bonds yielding 
directly E$_{Ga-P}$=-2.499 eV
a smaller value than in the clusters with pentagons. 
However, as it can be 
seen in Table II, it is the isomer with squares which has the lowest 
energy among the two  Ga$_{12}$P$_{12}$.  The 1.9 eV energy 
difference between them is much less than the 9 eV found for 
B$_{12}$N$_{12}$\cite{Toftlund}. This is  most probably due to the less ionic 
character of the GaP bonds. The difference in electronegativity of Ga 
and P is in fact $\sim$ 0.4 against $\sim$ 1  for BN.  
The observed spontaneous formation in our simulations 
of a cage with the same topology of C$_{28}$ shows a possible evolution 
pattern from ionized bulk fragments to classical fullerene cages formed 
by pentagons and hexagons. 

In summary, we have shown, by means of ab-initio Car-Parrinello 
Molecular Dynamics, that small GaP fullerenes  have 
highly symmetric structures, closed  electronic shells and 
large HOMO-LUMO gaps and cohesive energy. 
These clusters are thermally stable and remain in the same structure 
also when ionized. These findings together with 
the observed spontaneous formation  in our simulations
of a 28 atoms cage with the same symmetry of C$_{28}$  
support the possible existence of GaP fullerenes.
We hope that our work will stimulate experimental groups to widen their search 
for hetero-fullerenes also  to  III-V compound semiconductors.

{\bf Acknowledgments}: V.T. acknowledges financial support of the Research 
Institute for Materials (RIM) and of the Institute of Theoretical Physics of 
the University of Nijmegen  where this work was carried out. 
Calculations have been 
performed at the Stichting Academisch Rekencentrum Amsterdam (SARA) with 
a grant of the Stichting Nationale Computer Faciliteiten (NCF). 
We thank P. Giannozzi for providing relevant data on the pseudopotentials.
We are very grateful to A. Janner, R.A. de Groot and J.C. Maan for  
critical reading of the manuscript.

%\end{multicols}

\newpage

%-------------------------------------------------------------------
\begin{table}
\begin{tabular}{ccccc|ccccc}
\multicolumn{5}{c|}{Ga$_{16}$P$_{12}$ (T$_d$)}&\multicolumn{5}{c}{Ga$_{12}$P$_{16}$ (T$_d$)}\\
\hline
4 Ga  & 3m & xxx & x=-1.96 & r=3.39~~&   4 P   & 3m & xxx & x=-2.37 & r=4.11~~  \\
12 Ga &  m & xxz & x= 0.83 & r=3.50~~&   12 P  &  m & xxz & x= 0.81 & r=3.92~~ \\
      &    &     & z= 3.30 &        &         &    &     & z= 3.75 &        \\
12 P  &  m & xxz & x=2.87  & r=4.06~~&   12 Ga &  m & xxz & x= 2.33 &r=3.30~~ \\  
      &    &     & z=-0.03 &       ~~&         &    &     & z=-0.12&    \\
\end{tabular}
%\end{table}

%\begin{table}
\begin{tabular}{ccccc|ccccc|ccccc}
\multicolumn{5}{c|}{Ga$_{12}$P$_{8}$ (T$_h$)}&\multicolumn{5}{c|}{Ga$_{8}$P$_{12}$ (T$_h$)}&\multicolumn{5}{c}{Ga$_{12}$P$_{12}$ square (T$_h$)}\\
\hline
8 P   & 3 & xxx & x=2.05 & r=3.55~~& 8  Ga & 3 & xxx & x=1.54 & r=2.67~~& 12 P  & i  & 0yz &  y=3.36 & r=3.78~~\\
      &   &     &        &       &       &   &     &        &       &       &    &     &  z=1.74 &       \\
12 Ga & i & 0yz & y=2.61 & r=2.87~~& 12 P  & i & 0yz & y=3.18 & r=3.37~~& 12 Ga & i  & 0yz &  y=2.81 & r=3.13~~\\
      &   &     & z=1.19 &       &       &   &     & z=1.13 &       &       &    &     &  z=1.39 &       \\
\end{tabular}
\vspace{0.5cm}
\caption{Point symmetry and atomic positions of the clusters with symmetry T$_d$ and T$_h$. 
Values of the parameters and of the radial  distance r from the center of the cluster are 
in Angstrom. 
The positions of the minimum energy structure 
vary less than 0.01 \AA~~ from the given values.} 
\end{table}

\newpage

\begin{table}
\begin{tabular}{ccccc}
                        &       &          &         &	    \\
cluster                 & HOMO-LUMO gap   &E$_{cohesive}$ &  E$_{fit}$ & error \\
                        & (eV)  & (Hartree)     & (Hartree) &  (\%) \\
\hline						
Ga$_{12}$P$_{8}$	&1.28	&-2.5374& -2.5370&-0.01\\
Ga$_{8}$P$_{12}$	&2.14	&-2.8161& -2.8052&-0.39\\
Ga$_{16}$P$_{12}$	&1.09	&-3.6876& -3.6806&-0.19\\
Ga$_{12}$P$_{16}$	&1.55	&-3.9452& -3.9488&0.09\\
Ga$_{10}$P$_{10}$	&1.03	&-2.6632& -2.6711&0.30\\
Ga$_{12}$P$_{12}$(pentagon)&1.24&-3.2362& -3.2429&0.21\\
Ga$_{12}$P$_{12}$(square)& 1.86	&-3.3060 & -3.4308&3.78\\
\hline
\end{tabular}
\vspace{0.5cm}
\caption{HOMO-LUMO gap and cohesive energy of the clusters studied. 
For reference, the indirect energy gap of bulk GaP within the Local 
Density Approximation is 1.62 eV.  
The cohesive energy is obtained as the difference of the total energy minus 
the energy of the isolated pseudoatoms.
Within the same approximations these values are -2.15045 Hartree and 
-6.46295 Hartree for Ga and P, respectively. The cohesive energy per GaP pair in the bulk is found to be 0.2976 Hartree. In the last two columns we give
the values and the relative error  of the cohesive energy estimated from
the calculated bond energies E$_{Ga-P}$=
-2.593 eV,  E$_{Ga-Ga}$=-1.133 eV,  E$_{P-P}$=-2.349 eV (see text). 
}
\end{table}

%-------------------------------------------------------------------

\newpage
\begin{figure}

\caption{
Ga atoms are represented as large light balls, P atoms as small
dark balls.
Starting from a truncated bulk structure with tetrahedral symmetry,
Fig.1a shows a first step in the evolution towards structural energy 
minimization. The main rearrangement is the bonding of the peripheral Ga 
atoms between themselves, 12 atoms in pairs on the edges of the 
tetrahedron 
and four triplets on the vertices.  
In Fig.1b the central P atom breaks its bonds, 
followed by 12 Ga atoms (Fig1.c) leading to the formation of a
Ga$_{16}$P$_{12}$ fullerene cage. 
Fig.1d shows the equilibrium structure
of the neutral Ga$_{16}$P$_{12}$ cluster alone. 
Notice the non-planarity of the pentagons. 
The symmetry of the equilibrium configuration is T$_d$ (see Table I).
}
\end{figure}


\begin{references}
\bibitem{nota21} In Ref.\onlinecite{PRB2} we have noted that also
III-V clusters included in sodalite show the tendency to organize in 
hollow shells of cations and anions.
\bibitem{nota} 
The neutral Ga$_{28}$ P$_{13}$ cluster with 149 electrons has one unpaired 
electron in the uppermost occupied state. 
Just below this state,  the electronic structure
presents one 3--fold  and one 1--fold degenerate state. 
Interestingly,  the 28 atoms cage forms 
spontaneously whenever we have completely filled states, namely for 
148, 142 and 140 electrons, which would describe an ionized bulk-like fragment.
\end{references}
\end{document}